\documentclass[preprint,showpacs,preprintnumbers,amsmath,amssymb,
superscriptaddress,unsortedaddress]{revtex4}
\usepackage{graphicx}% Include figure files
\usepackage{dcolumn}% Align table columns on decimal point
\usepackage{bm}% bold math

\begin{document} 

%\preprint{APS/123-QED}

\title{Quantum walks on cycles}

\author{Ma{\l}gorzata Bednarska}
%\email{mbed@amu.edu.pl}
\affiliation{Faculty of Mathematics 
and Computer Science,
 Adam Mickiewicz University,
Umultowska 87, 61-614 Pozna\'{n}, Poland.
}
\author{Andrzej Grudka}%
%\email{agie@amu.edu.pl}
\affiliation{Faculty of Physics,
 Adam Mickiewicz University,
Umultowska 85, 61-614 Pozna\'{n}, Poland.
}
\author{Pawe{\l} Kurzy\'nski}
%\email{@amu.edu.pl}
\affiliation{Faculty of Physics,
 Adam Mickiewicz University,
Umultowska 85, 61-614 Pozna\'{n}, Poland.
}
\author{Tomasz {\L}uczak}
%\email{tomasz@amu.edu.pl}
\affiliation{Faculty of Mathematics 
and Computer Science,
 Adam Mickiewicz University,
Umultowska 87, 61-614 Pozna\'{n}, Poland.
}
\author{Antoni W\'ojcik}
%\email{antwoj@amu.edu.pl}
\affiliation{Faculty of Physics,
 Adam Mickiewicz University,
Umultowska 85, 61-614 Pozna\'{n}, Poland.
}

%\date{April 15, 2003}% It is always \today, today,

\begin{abstract}
We consider asymptotic  behaviour of a Hadamard walk on a cycle. 
For a walk which starts with a state in which all the probability is 
concentrated on one node, 
we find the explicit formula for the limiting distribution and discuss 
its asymptotic behaviour when the length of the cycle tends to infinity.
We also demonstrate that for a carefully chosen initial state,
the limiting distribution of a quantum walk on cycle can lie 
further away from the uniform distribution than its initial state.
\end{abstract}

\maketitle

\section{\label{sec:intro}
Introduction}

Celebrated results of Shor~\cite{Shor} and Grover~\cite{Grover} started a quest for
algorithms based on quantum mechanics which can surpass corresponding classical algorithms. 
As many classical algorithms employ properties of  random 
walks on graphs, several groups of researchers 
have begun to study properties of quantum analogues of 
classical random walks (for a survey see~\cite{Kempe1}).
Two models of quantum random walks have been proposed by Aharonov {\it et al.\/}~\cite{Aharonov} 
and Farhi and Gutmann~\cite{Farhi}; in the paper we consider 
only a discrete quantum walk as defined in~\cite{Aharonov}.
The behavior of quantum random walks have been shown to differ greatly from  
their classical counterparts. Ambainis {\it et al.\/}~\cite{Amb} proved that the spreading time in 
the quantum walk on line scales linearly with the number of steps, 
while Aharonov {\it et al.\/}~\cite{Aharonov}
showed that the mixing time for a walk on a cycle grows linearly with the cycle length. 
Even more  spectacular exponential speed up was discovered by  Kempe {\it et al.\/}~\cite{Kempe2} 
who studied  a quantum random walk on hypercube; this result led to the construction of the first 
quantum algorithm based on a random walk by Shenvi {\it et al.\/}~\cite{Shenvi}. 
The fact that in quantum walks on graphs the probability 
function spreads out much faster than in classical case 
is  not the only factor which can be explored  in designing new  quantum  
algorithms.  In~\cite{Aharonov} the authors remarked that `one 
may try to use quantum walks which converge to limiting distribution which are different than 
those of the corresponding classical walks'. In this note we follow this suggestion 
and study the limiting distribution of a random walk on a cycle; it turns out
that it depends on the length of the cycle in a somewhat surprising way.
We also give an example of a quantum walk in which the distance from the initial 
state to the uniform distribution is larger than the distance between the uniform
distribution and the initial state. 
Finally, we remark that recently Travaglione and Milburn~\cite{Tr}
and D\"ur {\it et al.}~\cite{D} have proposed a scheme of experiment which  realizes 
a quantum walk on a cycle.

\section{\label{sec:model}
Model}

We study a quantum random walk on a cycle with $d$ nodes. In the model of 
such a walk proposed in~\cite{Aharonov} nodes are represented by vectors   
$|v\rangle$, $v=0,1,\dots,d-1$, which form an orthonormal basis of the Hilbert space $H_V$. 
An auxiliary two-dimensional Hilbert space $H_A$  (coin space) is spanned by vectors $|s\rangle$, 
$s=0,1$. 
The initial state of the walk is a normalized vector 
\begin{equation}\label{eq0}
|\Psi_0\rangle=\sum_{s,v}\alpha_{sv}|s,v\rangle=\sum_{s,v}\alpha_{sv}|s\rangle|v\rangle
\end{equation} 
from the  tensor product $H=H_A\otimes H_V$. In a single step of the walk the 
state changes according to the equation 
\begin{equation}\label{eq1}
|\Psi_{n+1}\rangle=U|\Psi_n\rangle,
\end{equation}
where the operation $U=S(H\otimes I)$  first applies the Hadamard gate operator
$H=\sum_{s,s'}(-1)^{ss'}|s\rangle\langle s '|$ to the vector from $H_A$, and then    
shifts the state  by the operator 
\begin{equation}\label{eq2}
S= \sum_{s,v}|s\rangle\langle s |\otimes|v+2s-1 ({\rm mod}\  d)\rangle\langle v|\,.
\end{equation}

The probability distribution on the nodes of the cycle after 
the first $n$ steps of the walk is given by 
\begin{equation}\label{eq3}
p_n(v)=\sum_s|\langle s,v|\Psi_n\rangle|^2.
\end{equation}
However, as was observed by Aharonov {\it  at el.\/}~\cite{Aharonov}, for a fixed $v$, 
the probability  $p_n(v)$ is `quasi-periodic' as a function of $n$ and thus, typically,
it does not converge to a limit. Thus, instead of $p_n(v)$ the authors of~\cite{Aharonov} considered 
\begin{equation}\label{eq33}
\bar p_n(v)=\frac1n\sum_{i=1}^n p_n(v),
\end{equation}
and proved that for any initial state $|\Psi_0\rangle$ and every node $v=0,1,\dots,d-1$,
the sequence  $\bar p_n(v)$ converges to the limiting distribution 
\begin{equation}\label{eq5}
%\begin{aligned}
\pi(v)=\sum_{a,a'}\sum_{s}\Gamma_{aa'}|\langle\phi_a|\Psi_0\rangle\langle\Psi_0|\phi_{a'}\rangle
%\\&\times
\langle s ,v|\phi_a\rangle\langle\phi_{a'}|s,v\rangle,
%\end{aligned}
\end{equation}
where $c_a$ are the eigenvalues of $U$, $|\phi_a\rangle$ stand for 
the eigenvectors of $U$, and 
\begin{equation}\label{eq6}
\Gamma_{aa'}=\begin{cases}
1&\quad {\rm if}\quad c_a=c_{a'}\\
0&\quad {\rm if}\quad c_a\neq c_{a'}.
\end{cases}
\end{equation} 

In the  case of the Hadamard walk on cycle, for $j=0,1$
and $k=0,1,\dots,d-1$, we get 
\begin{equation}\label{eq7}
c_{jk}=\frac{1}{\sqrt{2}}\big( (-1)^k\sqrt{1+\cos^2(2\pi j/d)}-
i\sin(2\pi j/d) \big),
\end{equation}
and
\begin{equation}\label{eq8}
|\phi_{jk}\rangle=(a_{jk}|0\rangle+a_{jk}b_{jk}|1\rangle)\otimes\sum_v\omega_d^{jv}|v\rangle,
\end{equation}
where $\omega_d=e^{2\pi i/d}$, 
\begin{align}
a_{jk}&=1\Big/\sqrt{d(1+|b_{jk}|^2)},\label{eq9}\\
b_{jk}&=\omega_d^j\big( (-1)^k\sqrt{1+\cos^2(2\pi j/d)}-
\cos(2\pi j/d) \big).\label{eq10}
\end{align}
Note that $c_{j0}\neq c_{j'1}$ so $\Gamma_{jk,j'k'}=\gamma_{jj'}\delta_{kk'}$, 
where $\delta_{kk'}$  is the Kronecker delta. 
Thus, (\ref{eq5}) becomes
\begin{equation}\label{eq5b}
%\begin{aligned}
\pi(v)=\sum_{j,j'}\sum_{k}\gamma_{jj}|\langle\phi_{jk}|\Psi_0\rangle\langle\Psi_0|\phi_{j'k}\rangle
%\\&\times 
A_{jj'k}\omega_d^{v(j-j')},
%\end{aligned}
\end{equation}
where  $A_{jj'k}=a_{jk}a_{j'k}(1+b_{jk}b^*_{j'k})$. If $d$ is odd, then all eigenvalues are 
distinct, $\gamma_{jj'}=\delta_{jj'}$, and $A_{jjk}=1/d$;
consequently, $\pi(v)=(1/d)\sum_j\sum_k|\langle\phi_{jk}|\Psi_0\rangle|^2=1/d$. 
It comes as no surprise,
since as was proved in  Aharonov {\it et al.}~\cite{Aharonov} 
the limiting distribution $\pi$ is  always uniform in a non-degenerate case.
Thus we concentrate on  more interesting case  of even $d$.
Then,  because of symmetries $c_{d/2-j,k}=c_{j,k}$
and $c_{d/2+j,k}=c_{jk}^*=c_{d-j,k}$, the coefficient  $\gamma_{jj'}$
does not vanish when one of the following conditions hold:
\begin{enumerate}
\item $j=j'$;
\item $j=0$, $j'=d/2$;
\item $j=t$, $j'=d/2-t$, for $t=1,2,\dots,t_{\rm max}$;
\item $j=d-t$, $j'=d/2+t$, for $t=1,2,\dots,t_{\rm max}$;
\end{enumerate} 
where here and below $t_{\rm max}=\lfloor (d-2)/4\rfloor$.

\section{\label{sec:result}Result and discussion}

We study in detail the limiting distribution $\pi$ for a quantum 
walk for which starts with a state in which with probability one 
a particle is at a node $v_0$; more specifically we set 
\begin{equation}\label{e13}
|\Psi_0\rangle=\frac{1}{\sqrt 2}\big(|0,v_0\rangle+i|1,v_0\rangle\big).
\end{equation}

Then $\langle\phi_{jk}|\Psi_0\rangle =g_{jk}\omega_d^{-v_0j}$, 
where $g_{jk}=a_{jk}(1+ib^*_{jk})\big/\sqrt 2$, and 
\begin{equation}\label{e14}
\pi(v)=\sum_{j,j'}\sum_k\gamma_{jj'}f(j,j',k)\omega_d^{(v-v_0)(j-j')},
\end{equation}
with $f(j,j',k)=g_{jk}g^*_{j'k}A_{jj'k}$. 
Since $\gamma_{jj'}=0$ except of the four  cases (i)--(iv) described above, 
we get 
\begin{equation}\label{e15}
\pi(v)=F+(-1)^\Delta 2 {\rm Re} (F_0)+(-1)^\Delta{\rm Re}\Big(\sum_{i=1}^{t_{\rm max}}F_t\Big),
\end{equation}
where by 
\begin{equation}\label{e15a}
\Delta=\Delta(v)=\min\{|v-v_0|,d-|v-v_0|\}
\end{equation}
denotes the distance between nodes $v_0$ and $v$, and
\begin{equation}\label{e16}
\begin{aligned}
F&=\sum_{j,k} f(j,j,k),\\
F_0&=\sum_k f(0,d/2,k),\\
F_t&=\sum_k f(t,d/2-t,k)+f(d/2+t,d-t,k).
\end{aligned}
\end{equation}
After some elementary but not very exciting 
calculations (\ref{e16}) reduces to 
\begin{equation}\label{e16a}
\begin{aligned}
F&=\frac{1}{d},\\
F_0&=0,\\
F_t&=\frac{2}{d^2}\frac{\sin^2(2\pi t/d)}{1+\cos^2(2\pi t/d)}.
\end{aligned}
\end{equation}
We remark that $F_t=0$ whenever $t_{\rm max}=0$, i.e., for $d=2$ and $d=4$.
Hence the limiting distribution for even cycles of sizes two and four 
are uniform, which has also been observed by Travaglione and Milburn~\cite{Tr},
who analyzed a quantum walk on cycle of length four step by step.  
However, for $d\ge 6$ we have 
\begin{equation}\label{eqq}
\pi(v)=(1+\Pi(v))/d, 
\end{equation}
with the `correction' term 
$\Pi(v)=(-1)^\Delta 4S/d$, where 
\begin{equation}\label{e21}
S=\sum_{t=1}^{t_{\rm max}}\cos(4\pi t\Delta/d)\Big(\frac{2}{1+\cos^2(2\pi t/d)}-1 \Big).
\end{equation}
Setting $z=3-2\sqrt 2\sim 0.17157\dots$, one can write (\ref{e21}) as
\begin{equation}\label{e22}
\begin{aligned}
S=\Big(\frac{8z}{1-z^2}-1\Big)\sum_{t=1}^{t_{\rm max}}\cos(4\pi\Delta t/d)
&+\frac{8z}{1-z^2}\sum_{m=1}^{\infty}(-z)^m  
\sum_{t=1}^{t_{\rm max}}  \cos(4\pi(\Delta+m)t/d)\\
&+\frac{8z}{1-z^2}\sum_{m=1}^{\infty}(-z)^m  
\sum_{t=1}^{t_{\rm max}}  \cos(4\pi(\Delta-m)t/d),
\end{aligned}
\end{equation}  
which, in turn, transforms to 
\begin{equation}\label{e23}
\begin{aligned}
S=-\frac{d}{4}(\delta_{\Delta,0}+\delta_{\Delta,d/2})
+&(-1)^\Delta \frac d4\,\frac{8z}{1-z^2}\,\frac{z^{-\Delta}(-z)^{d/2}+z^\Delta}
{1-(-z)^{d/2}}\\
&+\frac{1+(-1)^\Delta \xi}{2}-\frac{4z}{(1-z)^2}-\frac{4z(-1)^\Delta\xi}{(1+z)^2},
\end{aligned}
\end{equation}  
where $\xi=\big(1+(-1)^{d/2}\big)\big/2$, i.e., $\xi=1$ when $d/2$ is even,
and $\xi=0$ if $d/2$ is odd. Thus, we arrive at 
%the following formula for $\Pi(v)$
\begin{equation}\label{e24}
\begin{aligned}
\Pi(v)=(-1)^{\Delta+1}(\delta_{\Delta,0}+\delta_{\Delta,d/2})
+&\frac{8z}{1-z^2}\,\frac{z^{-\Delta}(-z)^{d/2}+z^\Delta}
{1-(-z)^{d/2}}\\
&+\frac{4}{d}\Big(\frac{\xi+(-1)^\Delta}{2}-\frac{4(-1)^\Delta z}{(1-z)^2}
-\frac{4z\xi}{(1+z)^2}\Big).
\end{aligned}
\end{equation}

\begin{figure}
\includegraphics{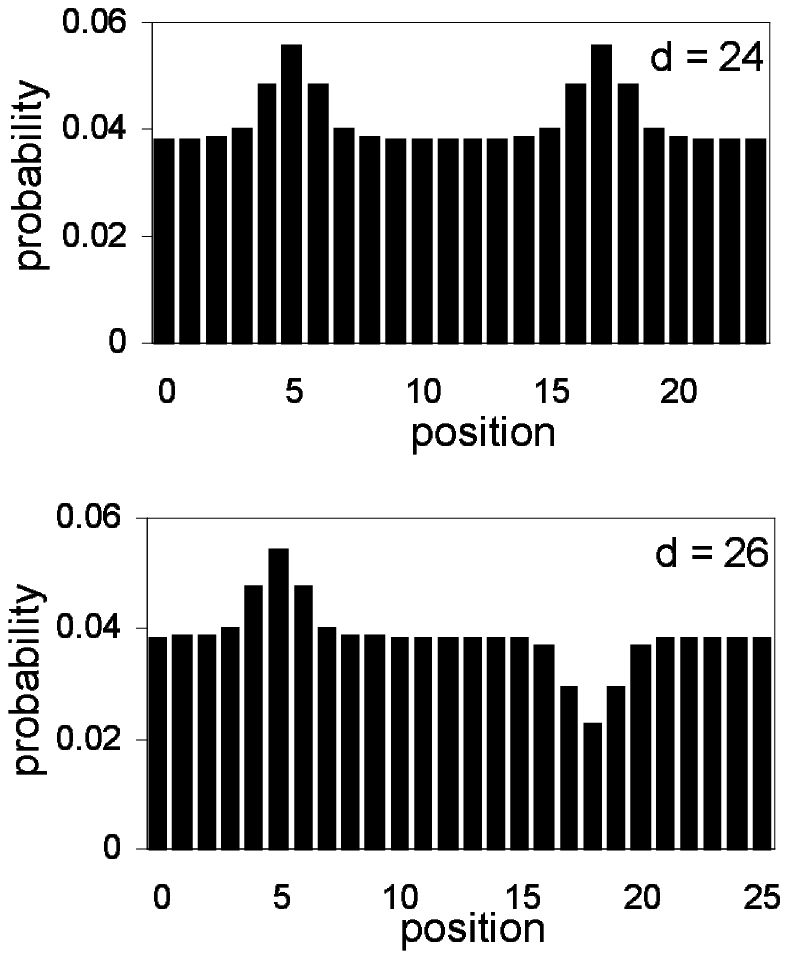}% Here is how to import EPS art
\caption{\label{f1} The limiting distributions for the cases of $d=24$   
and $d=26$. The walk starts with the state where with probability one
a particle is at the node $v_0=5$.}
\end{figure}

\begin{figure}
%\scalebox{.7}
{\includegraphics{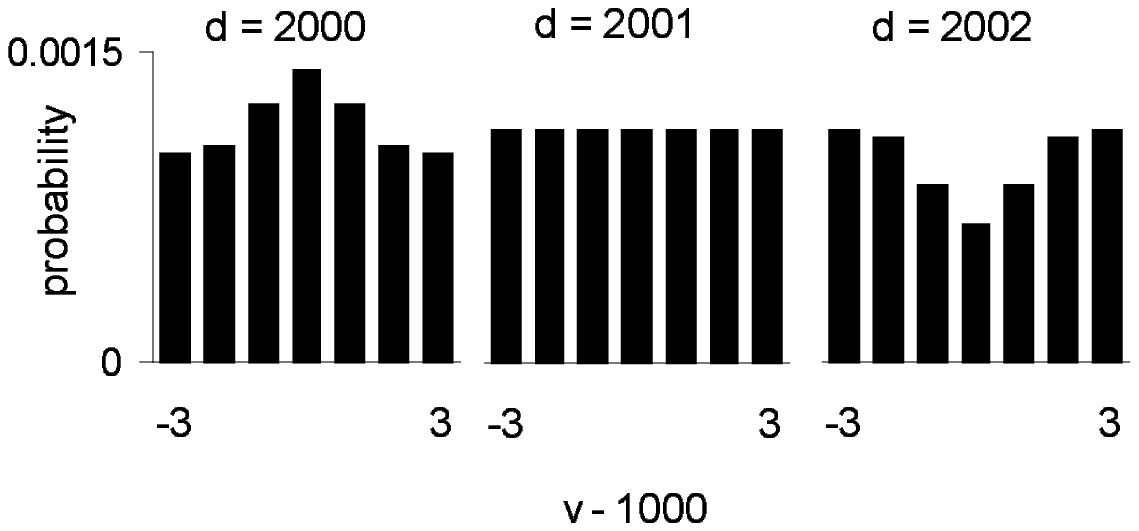}}% Here is how to import EPS art
\caption{\label{f3} The limiting distribution for cycles of lengths $2000$, 
$2001$ and $2002$. In each case the walk starts with a state in which 
with probability one a particle is at  $v_0=0$.}
\end{figure}

In Figure~\ref{f1} we pictured the resulting limiting distributions for 
$d=24$  and  $d=26$. It is easy to see they are almost uniform except 
for the nodes which lie next to  the initially populated node $v_0$   and 
the opposite node $\hat v_0=v_0+d/2$ \mbox{$({\rm mod}\ d)$}. 
Indeed, as $d\to\infty$, 
the last term of (\ref{e24}) vanishes and
\begin{equation}\label{e24b}
\bar\Pi(v)=\lim_{d\to\infty}\Pi(v)=\eta(\Delta)-(-1)^\xi\eta(\Delta'),
\end{equation}
where
\begin{equation}\label{e24c}
\eta(x)=\frac{8z^{1+x}}{1-z^2}-\delta_{x0},
\end{equation}
and $\Delta'$ is the distance between nodes $v$ and $\hat v_0$.
The equation~(\ref{e24b}) shows that the correction term $\bar\Pi(v)$ 
is significant only for $v$ which lie close to either $v_0$ or 
$\hat v_0$ and decreases exponentially with the distance between 
$v$ and $v_0$ and $\hat v_0$.
Note that the shapes of the cusps  near $v_0$ 
and $\hat V_0$ does not depend very much on $d$ for large $d$
(except of the scaling factor $1/d$).
However, if $d/2$ is odd then the limiting distribution $\pi(v)$ has a minimum at $\hat v_0$,
while if $d/2$ is even the distribution has a peak  $\hat v_0$, virtually 
identical with that which appear at $v_0$ (see Figures~\ref{f1} and \ref{f3}).
The fact that such a  local behaviour of the limiting distribution 
$\pi$ depends so strongly on `global' properties of space, as the parity of $d/2$,
is somewhat surprising. We hope that this and/or analogous phenomena can be used
in constructing efficient quantum algorithms.

%\begin{figure}
%\includegraphics{fig2}% Here is how to import EPS art
%\caption{\label{f2} Function  $\eta(\Delta)$  defined in~\ref{e24c}.}
%\end{figure}

\begin{figure}
\includegraphics{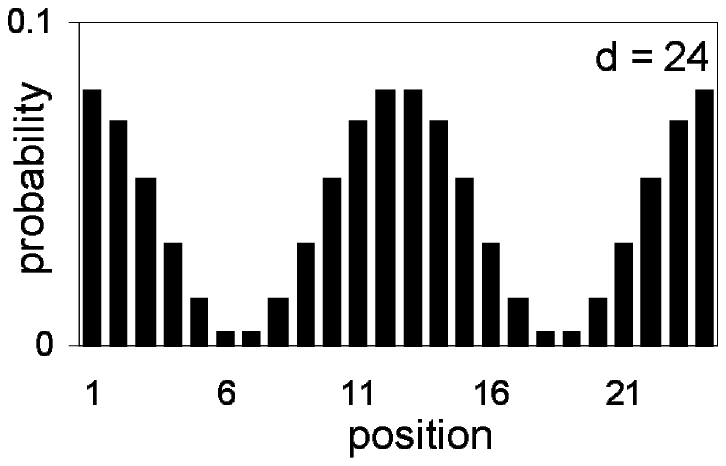}% 
\caption{\label{f4} An example of a non-uniform invariant probability distribution for
an initial state of the form  $|\Psi_0=\frac{1}{\sqrt2}(|\phi_{5,0}\rangle + |\phi_{7,0}\rangle  )$.}
\end{figure}

It is easy to see that, despite of a narrow cusps near nodes $v_0$ and $\hat v_0$,
the total variation distance $d_{TV}$ between the limiting distribution given by~(\ref{eqq}) 
and (\ref{e24}) and the uniform distribution tends to 0 as $d\to\infty$. 
However it is not hard as well to construct `highly non-uniform' 
distributions which remain invariant during the walk: one can simply take as 
the initial state a  superposition of two degenerated eigenvectors
(the distribution  corresponding to a single eigenvector is 
always a uniform one). 
An example of such an invariant nonuniform 
probability distribution is presented on Figure~\ref{f4}.   

\begin{figure}
\includegraphics{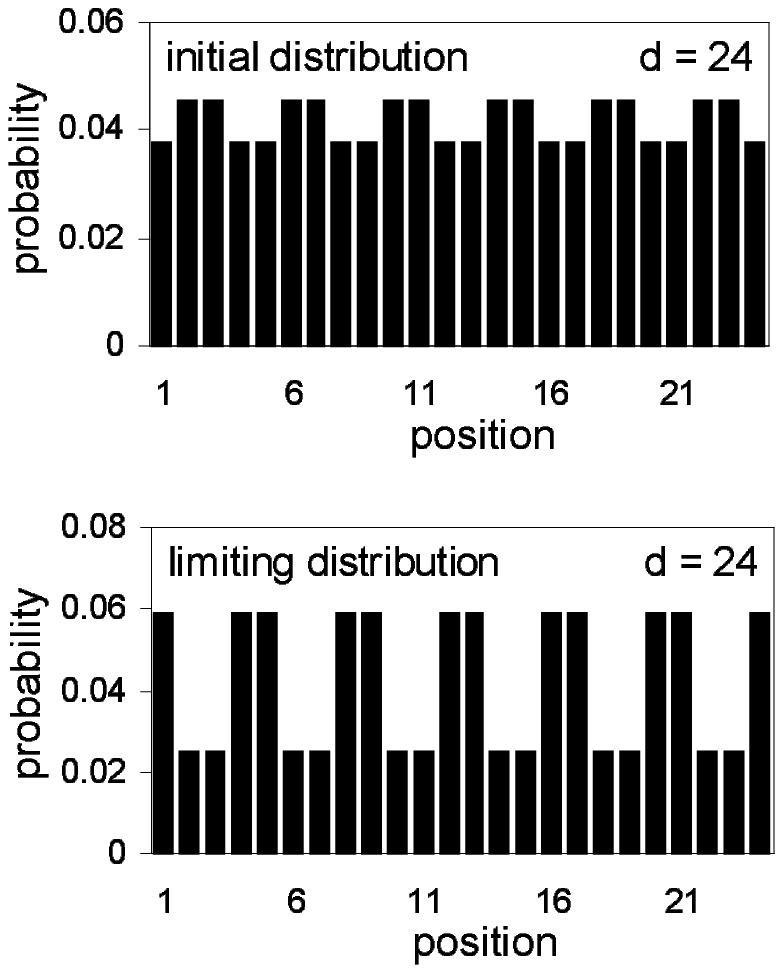}% 
\caption{\label{f5} The initial and limiting distributions for the initial state
$|\Psi_0\rangle =\frac{1}{2}(|\phi_{3,0}\rangle 
+|\phi_{9,0}\rangle - |\phi_{15,0}\rangle  - |\phi_{21,0}\rangle  )$.}
\end{figure}

With a slightly more work one can construct the initial states 
which indicate differences between quantum and classical walks on 
cycles even more distinctively. One such example is shown on Figure~\ref{f5}.
Here the total variation distance from the uniform distribution,
defined as 
\begin{equation}
\frac{1}{2}\sum_{v=0}^{d-1}\Big|p(v)-\frac{1}{d}\Big|
\end{equation}
is equal $0.046$ for the initial state, while for the
limiting distribution it grows to $0.204$. 
\vskip0.5truecm

\section{\label{sec:concl}
Conclusion}

We study the properties of a Hadamard quantum walk on a cycle with $d$ nodes. In 
the case of a walk starting from a single node we give an explicit formula 
for the limiting distribution and show that it is very sensitive to 
the arithmetic properties $d$. We hope that this or an analogous mode of 
behaviour  can be used in  construction of efficient quantum algorithms. 

Furthermore, we present an example of a quantum walk on a cycle,
for which the total variation distance between the initial distribution 
and the uniform  distribution is much smaller than the distance 
between limiting distribution and the uniform one.

%\vskip1truecm

\begin{acknowledgments}
We wish to thank the State Committee for Scientific Research 
(KBN) for its support: M.B.\ and T.\L.\ were supported by grant 2 P03A 016 23; A.G.\ and A.W.\ 
by grant 0~T00A~003~23. 
\end{acknowledgments}

%\bibliography{bezfaz}% Produces the bibliography via BibTeX.

\end{document}